\documentclass{midl} 

\usepackage{mwe} 

\jmlrproceedings{MIDL 2020}{Medical Imaging with Deep Learning 2020}
\jmlrvolume{}
\jmlrpages{}
\jmlryear{}
\jmlrworkshop{MIDL 2020 -- Short Paper}
\editors{}

\title[Deep Learning for Motion Forecasting Using 4D OCT Data]{A Deep Learning Approach for Motion Forecasting Using 4D OCT Data}

\midlauthor{\Name{Marcel Bengs\midljointauthortext{Contributed equally}\nametag{$^{1}$}} \Email{marcel.bengs@tuhh.de}\\
\Name{Nils Gessert\midlotherjointauthor\nametag{$^{1}$}} \Email{nils.gessert@tuhh.de}\\
\Name{Alexander Schlaefer\nametag{$^{1}$}} \Email{schlaefer@tuhh.de}\\
\addr $^{1}$ Institute of Medical Technology, Hamburg University of Technology, Hamburg, Germany
}

\begin{document}

\maketitle


\begin{abstract}
Forecasting motion of a specific target object is a common problem for surgical interventions, e.g. for localization of a target region, guidance for surgical interventions, or motion compensation. Optical coherence tomography (OCT) is an imaging modality with a high spatial and temporal resolution. Recently, deep learning methods have shown promising performance for OCT-based motion estimation based on two volumetric images. We extend this approach and investigate whether using a time series of volumes enables motion forecasting. We propose 4D spatio-temporal deep learning for end-to-end motion forecasting and estimation using a stream of OCT volumes. We design and evaluate five different 3D and 4D deep learning methods using a tissue data set. Our best performing 4D method  achieves motion forecasting with an overall average correlation coefficient of 97.41\%, while also improving motion estimation performance by a factor of 2.5 compared to a previous 3D approach. 
\end{abstract}


\section{Introduction}

Forecasting of patient or surgery tool movements is a relevant problem for various medical procedures ranging from radiotherapy \cite{ren2007adaptive} to ophthalmic interventions \cite{kocaoglu2014adaptive}. Typically, motion tracking requires a fast imaging modality. Optical coherence tomography (OCT) is an imaging modality with a high spatial and temporal resolution, allowing for 4D real-time imaging \cite{wang2016heartbeat}. OCT has been
integrated into intraoperative microscopes \cite{lankenau2007combining}, and has also been considered for monitoring laser cochleostomy \cite{pau2008imaging}. Often, OCT applications operate on small-scale, delicate structures where motion can disrupt the workflow or even cause injury \cite{bergmeier2017workflow}, hence accurate motion tracking is particularly relevant, e.g., for field of view (FOV) adjustment during intraoperative imaging \cite{kraus2012motion} or for adjustment of surgery tools during automated interventions \cite{zhang2014optical}. Recently, deep learning methods have shown promising results for motion estimation using two OCT volumes \cite{gessert2019two, laves2019deep}, or for pose estimation of a marker object using a single OCT volume \cite{gessert2018deep}.  While these methods have inference times in the range of milliseconds \cite{gessert2019two}, performing the actual compensation or adjustment introduces a lag between the actual adjustment and the motion estimation. This can be problematic if fast and large motions occur. One approach to overcome this problem is motion \textit{forecasting} for predicting the future trajectory, which requires a time series, instead of just two volumes for estimating a motion vector \cite{gessert2019two}. Processing sequences of 3D volumes leads to the challenging problem of 4D deep learning, where immense computational and memory requirements make architecture design very difficult. In this work, we propose an end-to-end deep learning approach for motion estimation and forecasting using entire sequences of OCT volumes. So far, 4D deep learning has only been considered for few examples, e.g., in the context of functional magnetic resonance imaging \cite{bengs2019a} and computed tomography \cite{clark2019convolutional}. We evaluate several 4D deep learning methods using a tissue data set and demonstrate a new mixed 3D-4D deep learning approach.

\section{Methods}
\textbf{Deep Learning Methods.}
We formulate a supervised learning problem where we predict the current motion vector $\Delta s_{t_{n}}\in\mathbb{R}^{3} $ as well as future motion vectors $\Delta s_{t_{n+1}}\in\mathbb{R}^{3}$, $\Delta s_{t_{n+2}}\in\mathbb{R}^{3}$ of a region of interest (ROI), given a 4D OCT image sequence $x_{{t}}= [x_{t_{0}},x_{t_{1}},...,x_{t_{n}}]$ capturing the history of a trajectory. We employ a dense neural network \cite{huang2017densely} as a baseline, using 3 densenet blocks with 3 layers each, connected with average pooling layers. Before the final regression layer of the network with 9 outputs, we use global average pooling. Using this baseline network, we evaluate five methods for processing of the 4D data, shown in \figureref{fig:model}. First, we only consider the initial and the last volume of a sequence processed by an architecture that has been proposed for motion compensation \cite{gessert2019two}, where two volumes are processed  by a two-path CNN with shared weights. The outputs of the two-path CNN are concatenated into the feature dimension and then processed by our densenet baseline with 3D convolutions. (2-Path-CNN3D)
Second, we use the entire sequence of volumes and extend the two path approach to a multi path architecture with shared weights, while the number of paths is equal to the number of input volumes. (n-Path-CNN3D) 
Third, we consider the entire sequence of volumes and directly learn from both spatial and temporal dimensions by using 4D spatio-temporal convolutions. We employ three 4D convolutional layers followed by our densenet with 4D convolutions. (CNN4D)
Fourth, we use a mixed 3D-4D approach, by applying a multi-path architecture for individually processing each volume of a sequence. Then, we reassemble a temporal dimension by concatenating the outputs into a temporal dimension, and afterward we apply our baseline network with 4D spatio-temporal convolutions.  (n-Path-CNN4D) Fifth, we consider a gated recurrent neural network with convolutional gating operations \cite{Xingjian2015} in front of our 3D baseline CNN, similar to \cite{bengs2019a}. (GRU-CNN3D)

\begin{figure}
\floatconts
  {fig:model}
  {\caption{The five architectures we employ.}}
  {\includegraphics[width=0.88\linewidth]{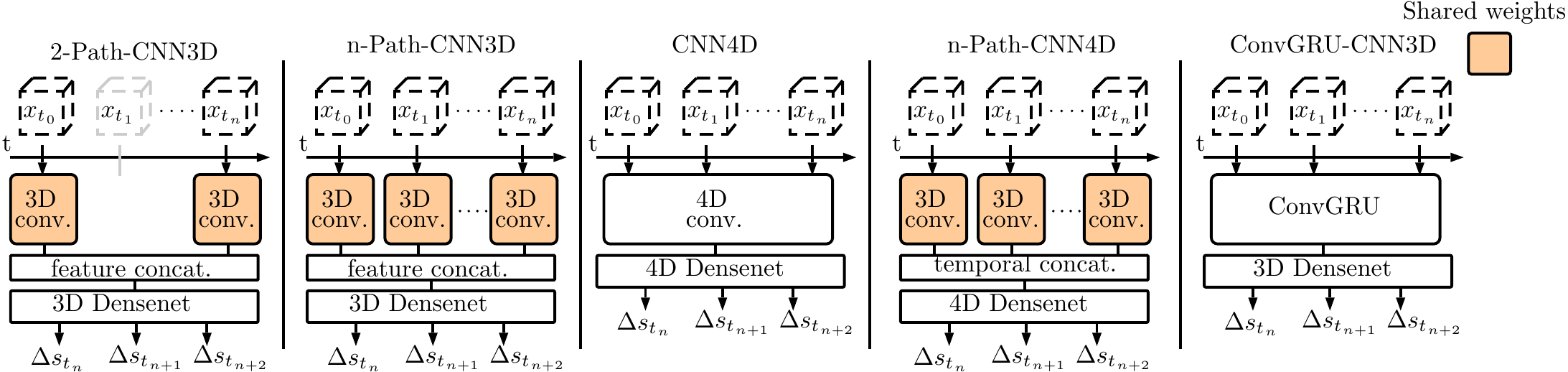}}
\end{figure}

\textbf{Data Set.}
We consider sequences of volumetric OCT images capturing motions of 40 different ROIs of a chicken breast sample, using a FOV of $5\,\mathrm{mm}\times5\mathrm{\,mm}\times3.5\mathrm{\,mm}$ and a volume size of $32\times32\times32$ pixels. We use the setup proposed by \citet{gessert2019two} with a swept-source OCT device (OMES, OptoRes) and a robot. This setup employs a scanning stage with mirror galvanometers allowing to shift the OCT's laser beam and thus the FOV, which can be utilized for automated data set generation and annotation. Using this setup we acquire sequences of volumetric OCT data $x_{{t}}=[x_{t_{0}},x_{t_{1}},x_{t_{2}},x_{t_{3}},x_{t_{4}}]$ for 100 different trajectories with corresponding labels $\Delta s_{{t}}= [\Delta s_{t_{4}},\Delta s_{t_{5}},\Delta s_{t_{6}}]$ for each of the 40 ROI. The ROIs have the same size as the OCT's FOV, hence at $t_{0}$ the FOV complete overlaps with a ROI. The label $\Delta s_{t_{4}}$ refers to the relative three dimensional motion vector between the initial position of a ROI at $t_{0}$ and the current one at $t_{4}$, while $\Delta s_{t_{5}}$ and $\Delta s_{t_{6}}$ are future relative motion vectors. For data generation, we consider various smooth curved trajectories with different motion magnitudes generated with spline functions. To evaluate our models on previously unseen ROIs, we randomly  choose  five independent  ROIs  for  testing  and  validating and use the remaining 30 ROIs for training. We optimize our networks for 400 epochs using Adam.

\section{Results and Discussion}

\begin{table}
 {\caption{Results for motion estimation ($\Delta s_{t_{n}}$) and  forecasting ($\Delta s_{t_{n+1}}$, $\Delta s_{t_{n+2}}$). Errors are given in mm. } \label{tab:network-details}}%
\centering
  {\begin{tabular}{llllll}
  &  MAE $\Delta s_{t_{n}}$ & MAE $\Delta s_{t_{n+1}}$ & MAE $\Delta s_{t_{n+2}}$  & aCC ($\%$) & Inf. Time (ms)\\ \hline
  2-Path-CNN3D & $0.40 \pm0.48$ & $0.48\pm0.61$   &$ 0.57\pm0.73$ & 77.49 & $3.61\pm0.37$  \\
  n-Path-CNN3D & $0.25\pm0.29$ &$ 0.31\pm0.36$ & $0.37\pm0.44$ & 92.27 & $4.51\pm0.17$  \\
  CNN4D  &$ 0.19 \pm0.21$ &$ 0.23\pm0.26$ & $0.28\pm0.31$ & 96.44
  & $12.52\pm0.16$  \\
  n-Path-CNN4D &  $0.16\pm0.18$ & $0.19\pm0.22$ &$ 0.23\pm0.27$ & 97.41 & $12.00\pm0.12$ \\
  GRU-CNN3D  &$ 0.20 \pm0.21$ &$ 0.25\pm0.26$ & $0.31\pm0.32$ & 95.49
  & $12.76\pm0.17$  \\
  
\hline
  \end{tabular}}
\end{table}

We report the mean absolute error (MAE) for estimation and forecasting separately, also we consider the average correlation (aCC) coefficient for combined performance in Table \ref{tab:network-details}. As expected, extending a previous approach (2-Path-CNN3D) to the processing of sequences (n-Path-CNN3D) already improves motion estimation and motion forecasting performance. However, comparing the different models for processing of the 4D data, n-Path-CNN3D performs worse. This indicates that temporal processing using the channel dimension is challenging, as already shown in the natural image domain \cite{tran2015learning}. 
While learning from the 4D data with CNN4D and GRU-CNN3D performs well, combining the multi path approach with a 4D architecture increases performance even further, suggesting an effective spatial preprocessing. Moreover, there are no optimized, native 4D convolution operations available so far and yet our computational expensive 4D models still achieve competitive motion estimates with up to 83 Hz. Overall, we demonstrate that 4D deep learning methods enable high accuracy motion forecasting based on time series of volumetric data.

\midlacknowledgments{This work was partially funded by Forschungszentrum Medizintechnik Hamburg (grants 04fmthh16).}

\bibliography{midl-samplebibliography}

\begin{thebibliography}{16}
\providecommand{\natexlab}[1]{#1}
\providecommand{\url}[1]{\texttt{#1}}
\expandafter\ifx\csname urlstyle\endcsname\relax
  \providecommand{\doi}[1]{doi: #1}\else
  \providecommand{\doi}{doi: \begingroup \urlstyle{rm}\Url}\fi

\bibitem[Bengs et~al.(2019)Bengs, Gessert, and Schlaefer]{bengs2019a}
Marcel Bengs, Nils Gessert, and Alexander Schlaefer.
\newblock 4d spatio-temporal deep learning with 4d fmri data for autism
  spectrum disorder classification.
\newblock In \emph{International Conference on Medical Imaging with Deep
  Learning}, 2019.

\bibitem[Bergmeier et~al.(2017)Bergmeier, Fitzpatrick, Daentzer, Majdani,
  Ortmaier, and Kahrs]{bergmeier2017workflow}
Jan Bergmeier, J~Michael Fitzpatrick, Dorothea Daentzer, Omid Majdani, Tobias
  Ortmaier, and L{\"u}der~A Kahrs.
\newblock Workflow and simulation of image-to-physical registration of holes
  inside spongy bone.
\newblock \emph{International Journal of Computer Assisted Radiology and
  Surgery}, 12\penalty0 (8):\penalty0 1425--1437, 2017.

\bibitem[Clark and Badea(2019)]{clark2019convolutional}
DP~Clark and CT~Badea.
\newblock Convolutional regularization methods for 4d, x-ray ct reconstruction.
\newblock In \emph{Medical Imaging 2019: Physics of Medical Imaging}, volume
  10948, page 109482A. International Society for Optics and Photonics, 2019.

\bibitem[Gessert et~al.(2018)Gessert, Schl{\"u}ter, and
  Schlaefer]{gessert2018deep}
Nils Gessert, Matthias Schl{\"u}ter, and Alexander Schlaefer.
\newblock A deep learning approach for pose estimation from volumetric oct
  data.
\newblock \emph{Medical image analysis}, 46:\penalty0 162--179, 2018.

\bibitem[Gessert et~al.(2019)Gessert, Gromniak, Schl{\"u}ter, and
  Schlaefer]{gessert2019two}
Nils Gessert, Martin Gromniak, Matthias Schl{\"u}ter, and Alexander Schlaefer.
\newblock Two-path 3d cnns for calibration of system parameters for oct-based
  motion compensation.
\newblock In \emph{Medical Imaging 2019: Image-Guided Procedures, Robotic
  Interventions, and Modeling}, volume 10951, page 1095108. International
  Society for Optics and Photonics, 2019.

\bibitem[Huang et~al.(2017)Huang, Liu, Van Der~Maaten, and
  Weinberger]{huang2017densely}
Gao Huang, Zhuang Liu, Laurens Van Der~Maaten, and Kilian~Q Weinberger.
\newblock Densely connected convolutional networks.
\newblock In \emph{CVPR}, pages 4700--4708, 2017.

\bibitem[Kocaoglu et~al.(2014)Kocaoglu, Ferguson, Jonnal, Liu, Wang, Hammer,
  and Miller]{kocaoglu2014adaptive}
Omer~P Kocaoglu, R~Daniel Ferguson, Ravi~S Jonnal, Zhuolin Liu, Qiang Wang,
  Daniel~X Hammer, and Donald~T Miller.
\newblock Adaptive optics optical coherence tomography with dynamic retinal
  tracking.
\newblock \emph{Biomedical optics express}, 5\penalty0 (7):\penalty0
  2262--2284, 2014.

\bibitem[Kraus et~al.(2012)Kraus, Potsaid, Mayer, Bock, Baumann, Liu,
  Hornegger, and Fujimoto]{kraus2012motion}
Martin~F Kraus, Benjamin Potsaid, Markus~A Mayer, Ruediger Bock, Bernhard
  Baumann, Jonathan~J Liu, Joachim Hornegger, and James~G Fujimoto.
\newblock Motion correction in optical coherence tomography volumes on a per
  a-scan basis using orthogonal scan patterns.
\newblock \emph{Biomedical optics express}, 3\penalty0 (6):\penalty0
  1182--1199, 2012.

\bibitem[Lankenau et~al.(2007)Lankenau, Klinger, Winter, Malik, M{\"u}ller,
  Oelckers, Pau, Just, and H{\"u}ttmann]{lankenau2007combining}
Eva Lankenau, David Klinger, Christian Winter, Asim Malik, Heike~Hedwig
  M{\"u}ller, Stefan Oelckers, Hans-Wilhelm Pau, Timo Just, and Gereon
  H{\"u}ttmann.
\newblock Combining optical coherence tomography (oct) with an operating
  microscope.
\newblock In \emph{Advances in medical engineering}, pages 343--348. Springer,
  2007.

\bibitem[Laves et~al.(2019)Laves, Ihler, Kahrs, and Ortmaier]{laves2019deep}
Max-Heinrich Laves, Sontje Ihler, L{\"u}der~A Kahrs, and Tobias Ortmaier.
\newblock Deep-learning-based 2.5 d flow field estimation for maximum intensity
  projections of 4d optical coherence tomography.
\newblock In \emph{Medical Imaging 2019: Image-Guided Procedures, Robotic
  Interventions, and Modeling}, volume 10951, page 109510R. International
  Society for Optics and Photonics, 2019.

\bibitem[Pau et~al.(2008)Pau, Lankenau, Just, and H{\"u}ttmann]{pau2008imaging}
HW~Pau, E~Lankenau, T~Just, and G~H{\"u}ttmann.
\newblock Imaging of cochlear structures by optical coherence tomography (oct).
  temporal bone experiments for an oct-guided cochleostomy technique.
\newblock \emph{Laryngo-rhino-otologie}, 87\penalty0 (9):\penalty0 641--646,
  2008.

\bibitem[Ren et~al.(2007)Ren, Nishioka, Shirato, and Berbeco]{ren2007adaptive}
Qing Ren, Seiko Nishioka, Hiroki Shirato, and Ross~I Berbeco.
\newblock Adaptive prediction of respiratory motion for motion compensation
  radiotherapy.
\newblock \emph{Physics in Medicine \& Biology}, 52\penalty0 (22):\penalty0
  6651, 2007.

\bibitem[Tran et~al.(2015)Tran, Bourdev, Fergus, Torresani, and
  Paluri]{tran2015learning}
Du~Tran, Lubomir Bourdev, Rob Fergus, Lorenzo Torresani, and Manohar Paluri.
\newblock Learning spatiotemporal features with 3d convolutional networks.
\newblock In \emph{ICCV}, pages 4489--4497, 2015.

\bibitem[Wang et~al.(2016)Wang, Pfeiffer, Regar, Wieser, van Beusekom, Lancee,
  Springeling, Krabbendam-Peters, van~der Steen, Huber,
  et~al.]{wang2016heartbeat}
Tianshi Wang, Tom Pfeiffer, Evelyn Regar, Wolfgang Wieser, Heleen van Beusekom,
  Charles~T Lancee, Geert Springeling, Ilona Krabbendam-Peters, Antonius~FW
  van~der Steen, Robert Huber, et~al.
\newblock Heartbeat oct and motion-free 3d in vivo coronary artery microscopy.
\newblock \emph{JACC: Cardiovascular Imaging}, 9\penalty0 (5):\penalty0
  622--623, 2016.

\bibitem[Xingjian et~al.(2015)Xingjian, Chen, Wang, Yeung, Wong, and
  Woo]{Xingjian2015}
S.~H.~I. Xingjian, Zhourong Chen, Hao Wang, Dit-Yan Yeung, Wai-Kin Wong, and
  Wang-chun Woo.
\newblock {Convolutional {LSTM} network: A machine learning approach for
  precipitation nowcasting}.
\newblock In \emph{Advances in Neural Information Processing Systems}, pages
  802--810, 2015.

\bibitem[Zhang et~al.(2014)Zhang, Pfeiffer, Weller, Wieser, Huber, Raczkowsky,
  Schipper, W{\"o}rn, and Klenzner]{zhang2014optical}
Yaokun Zhang, Tom Pfeiffer, Marcel Weller, Wolfgang Wieser, Robert Huber,
  J{\"o}rg Raczkowsky, J{\"o}rg Schipper, Heinz W{\"o}rn, and Thomas Klenzner.
\newblock Optical coherence tomography guided laser cochleostomy: Towards the
  accuracy on tens of micrometer scale.
\newblock \emph{BioMed research international}, 2014, 2014.

\end{thebibliography}

\end{document}